# Investing for Discovery and Sustainability in Astronomy in the 2020s

*Joan Najita, NOAO Chief Scientist*

As the next decade approaches, it is once again time for the US astronomical community to assess its investment priorities on the ground and in space in the coming decade. This report, created to aid NOAO in its planning for the 2020 Decadal Survey on Astronomy and Astrophysics, reviews the outcome of the previous Decadal Survey (Astro2010; Section 1); describes the themes that emerged from the 2018 NOAO Community planning workshop "NOAO Community Needs for Science in the 2020s" (Section 2); and based on the above, offers thoughts for the coming review (Section 3).

We find that a balanced set of investments in small- to large-scale initiatives is essential to a sustainable future, based on the experience of previous decades. While large facilities are the "value" investments that are guaranteed to produce compelling science and discoveries, smaller facilities are the "growth stocks" that are likely to deliver the biggest science bang per buck, sometimes with outsize returns. Investments in data-intensive missions also have benefits to society beyond the science they deliver. By training scientists who are well equipped to use their data science skills to solve problems in the public or private sector, astronomy can provide a valuable service to society by contributing to a data-capable workforce.

## 1. In Our Last Episode

The 2010 Decadal Survey, *New Worlds, New Horizons*, recommended a balance of large, medium, and small initiatives for both ground- and space-based astronomy. The highest priorities for ground-based astronomy (along with the estimated cost for the US federal share of the initiative, in FY2010 dollars) are shown in Table 1.

**Table 1. Recommendations for Ground-based Astronomy from the 2010 Decadal Survey**

| Initiative | Estimated Cost (US Federal Share, FY10 dollars) |
|---|---|
| **Large-Scale** | |
| **Large Synoptic Survey Telescope** (LSST)— "will transform observation of the variable universe and address broad questions ranging from the nature of dark energy to determining whether there are objects that may collide with Earth." | Construction: $421M<br>Operations: $28M/yr |
| **Mid-Scale Innovations Program** augmentation— "to respond | $93M to $200M over the decade |
| rapidly to scientific discovery and technical advances with new telescopes and instruments." | |
| **Giant Segmented Mirror Telescope** (GSMT)— "will revolutionize astronomy and spectroscopically complement JWST, ALMA, and LSST." | Construction: $257 to $350M<br>Operations: $9-14M/yr (for 25% US Federal share) |
| **Atmospheric Čerenkov Telescope Array** (ACTA)— "participation in an international telescope to study very high energy gamma rays. | Construction: $100M |
| **Medium-Scale** | |
| **CCAT** (formerly the Cornell-Caltech Atacama Telescope)— "a 25-meter wide-field submillimeter telescope that will complement ALMA by undertaking large-scale surveys of dust-enshrouded objects." | Construction: $37M<br>Operations: $7.5M |
| **Small-Scale (unranked)** | |
| Augmentation to Advanced Technologies and Instrumentation | + $5M/yr |
| Augmentation to Astronomy and Astrophysics Research Grants Program | + $8M/yr |
| Augmentation to Gemini international partnership | + $2M/yr |
| Augmentation to Telescope System Instrument Program | + $2.5M/yr |

(Rankings and cost in FY2010 dollars from the Executive Summary of *New Worlds, New Horizons*: https://www.nap.edu/read/12951/chapter/2#5)

Several of the above recommendations have made significant progress in this decade. LSST is on track for commissioning in mid-2020. Both US-led GSMTs (GMT, TMT) have partial funding (but no federal funding), and the NSF's midscale program has advanced a number of initiatives.

To date the MSIP program implemented by NSF has made approximately 24 grants totaling approximately $125M. Of these, approximately 8 (totaling approximately $40M) are related to ground-based optical astronomy. Most offer benefits to the US community in terms of publicly available data or open access observing time (Table 2).

**Table 2. MSIP Awards to Ground-based OIR Astronomy in the 2010s**

| Program | Award | Benefit to Community | NOAO Role? |
|---|---|---|---|
| **ZTF** construction, operation, and data management: develops skills and science important for the LSST era. | $11M | Open data from community surveys | Yes |
| **DES** data management at NCSA through the end of DES survey science operations | $7M | Single epoch images made available through the NOAO science archive, two public releases of data products | Yes |
| **PFS/Subaru** data reduction pipeline, | $5.5M | Public portal to the SuMIRe survey data | No |
| database and user interface for SuMIRe survey, a 600-night imaging and spectroscopic wide-field survey. | | (images and spectroscopy) | |
| **CHARA Array** | $4M | Open access through the NOAO TAC process | Yes |
| **Las Cumbres Observatory** | $3M | Open access time, available over 7 semesters, available through the NOAO TAC process | Yes |
| (MAPS) **MMT** AO exoPlanet characterzation System | $2.4M + $1.5M | Annual winter school for students | No |
| **Keck All Sky Precision AO** | $7M | All data publicly released | No |
| **LLAMAS**: an integral field spectrograph for the Magellan telescopes, designed for rapid and efficient spectroscopy of transients discovered by LSST | $2.4M | The instrument will be available to the US community through an open access agreement | ? |

(Reference: https://www.nsf.gov/awardsearch/advancedSearchResult?ProgEleCode=1257&BooleanElement=Any&BooleanRef=Any&ActiveAwards=true&#results)

As is apparent from the table, NOAO has played a significant role in the open access/open data aspect of MSIP over the last decade, i.e., through the allocation of MSIP-funded open time (through NOAO TAC) and the distribution of open data. The record of MSIP investments provides context for the 2020 Decadal Survey. They also provided background for the NOAO community meeting on Decadal Survey planning.

## 2. NOAO Community Meeting on Decadal Survey Planning

The NOAO community meeting "NOAO Community Needs for Science in the 2020s" (https://www.noao.edu/meetings/2020decadal/) was designed to assist the ground-based astronomical community in planning and preparing for the 2020 Decadal Survey. The workshop was also an opportunity for NOAO to receive valuable community input on the exciting scientific opportunities of the coming decade in areas in which NOAO can play a role in providing critical resources and/or areas that offer opportunities to strengthen the US ground-based OIR system. The science and resources described in the report from the Kavli Futures Symposium "Maximizing Science in the Era of LSST: A Community-based Study of Needed OIR Capabilities" provided background for this effort, as did community white papers submitted in advance of the workshop.

Held 20-21 February 2018 in Tucson, AZ, the workshop included high-level overviews of major science areas, reviews of current implementation plans for needed capabilities, as well as break out sessions on science topics, techniques, and capabilities in need of development (GSMTs, Wide-field Spectroscopy, High Angular Resolution). The agenda at the meeting website links to the archived presentations from all of the events

([https://www.noao.edu/meetings/2020decadal/#agenda](https://www.noao.edu/meetings/2020decadal/#agenda)).  Video streams of the plenary sessions enabled remote participation during the workshop. Highlights from the meeting are described below.

## Flagship Facilities

One of the take aways from meeting, as archived in the above materials, echoes the 2010 Decadal Survey: we expect that tremendous science will be enabled by the flagship facilities LSST ([talk by K. Bechtol](#)) and JWST ([talk by J. Lotz](#)) and the tools and follow-up resources needed to capitalize scientifically on the science opportunities they offer.

In particular LSST, which is on track to deliver the widest and deepest survey datasets, will usher in the astronomical "Big Data" era in a major way, cataloging more stars (17 billion), galaxies (20 billion), and solar system objects (5 million) than all previous astronomical surveys combined. The data quality and homogeneous processing of LSST data will enable new discoveries. However, the unprecedented data volume will present astronomers with the challenge of interacting with very large data sets and carrying out follow-up observations in an efficient way. Several presentations and community discussion focused on the strategies to meet these challenges (e.g., [talk by C. Schafer](#); [break out session on Science with Large Samples](#)).

## New Developments this Decade

While these highly anticipated flagship facilities featured prominently in the 2010 Decadal Survey, several new research directions have emerged in the intervening decade, engendering scientific enthusiasm and momentum.

The *Kepler Space Telescope*, launched less than 10 years ago (March 2009), revolutionized our view of exoplanets and attracted a huge fraction of our community, especially younger members, to this exciting new field of research. By January 2015, *Kepler* had found more than 1000 confirmed exoplanets in 440 star systems ([https://en.wikipedia.org/wiki/Kepler_(spacecraft)](https://en.wikipedia.org/wiki/Kepler_(spacecraft))).  A successor mission, TESS launched April 2018, will detect planets around brighter stars than *Kepler* (see talk by [C. Dressing](#)). As with *Kepler*, ground-based observations will be essential to maximize the science from *TESS*, and smaller aperture facilities will play a key role in vetting candidate planets and characterizing host stars.

In parallel, several developments have led to a new emphasis on stellar astronomy. These include the high precision, high cadence photometry of *Kepler*, the transformational astrometry from *Gaia*, Galaxy-wide spectroscopy with *APOGEE*, and a new generation of wide-field imaging studies (e.g., [*Dark Energy Survey*](#) and the [*Legacy Surveys*](#) carried out with Blanco/DECam, Mayall/Mosaic3, and Bok/90Prime).

Optical interferometry has also come of age in the last decade, moving beyond technical demonstrations and developing broader user communities. Facilities like CHARA are making fundamental measurements of stars (sizes, effective temperatures) that enable unique tests of stellar evolution, precise size measurements of transiting planets, and probes of stellar surface structure (limb darkening, convection and granulation; spots and magnetic activity), and possibly future detections of planet formation events (see talk by [G. Schaefer](#)).

Along these lines, several presentations (talks by [J. Fuller](#), [J. Kollmeier](#)) led to a sense that the 2020s would see a renaissance in stellar astronomy due to the "great convergence" of resources such as:
- high cadence photometry (*TESS, CHEOPS, PLATO; ZTF, Blackgem, ATLAS, LCO, ASAS-SN, PanSTARRS, LSST*)
- deep photometry (*LSST*)
- high accuracy astrometry (*Gaia*)
- wide-field spectroscopy (*SDSS-V, DESI, APOGEE, PFS, 4MOST, MOONS*)

These may lead to new discoveries about eclipsing binaries, pulsating stars, asteroseismology (stellar rotation, magnetic fields, internal convection), and other strange, unexpected phenomena (e.g., Tabby's star).

Another major milestone within the last decade is the arrival of the era of multi-messenger astronomy (talk by [D. Sand](#)). The combination of *LIGO, LSST, ZTF* and other experiments will drive time domain astronomy in the 2020s. Needed supporting capabilities are well known (see Matheson's [summary of the time domain breakout session](#)) and include alert brokers, target observation managers, and follow-up observing systems that provide access to workhorse instrumentation on a wide range of telescope apertures.

### On the Horizon

Capabilities that have featured prominently in previous decadal surveys also continue to be highly anticipated and highly valued.

Highly multiplexed, wide-field multi-object spectroscopy, which has been called out in both the 2000 and 2010 Decadal Surveys, is on the brink of deployment in the US with *PFS* on the 8-m Subaru telescope and the *Dark Energy Spectroscopic Instrument (DESI)* on the Kitt Peak 4-m Mayall telescope. While these facilities can address some important long-standing questions, many others appear to require a larger aperture facility. [Jeff Newman's presentation](#) summarized the options currently under investigation and quantified their ability to address marquee questions (e.g., those from the Kavli Futures Symposium on "Maximizing Science in the Era of LSST").

Giant Segmented Mirror Telescopes (GSMTs) have been highly ranked in both the 2010 and 2020 Decadal Surveys. Both US GSMTs (GMT and TMT) enable new science and will be extremely powerful when used independently or in concert to address major science questions. The latter is a current focus of the [US Extremely Large Telescope Program](#).

Both GSMTs have made terrific progress in the past decade, but the more rapid advance of the E-ELT relative to the US GSMTs is a concern with regard to US leadership in astronomy.

## Summary: Science Forecast for the 2020s

To summarize the NOAO community meeting in a different way, we can make the following science forecast for the coming decade, based on the presentations and discussions at the meeting:

**Astronomers and the public will continue their love affair with planets**. Driven by space missions and ground-based follow up, the 2020s will see continued growth and depth in our understanding of the exoplanet population. (The capabilities needed to achieve this include ground-based precision radial velocity spectroscopy, stellar spectroscopy, speckle imaging, longer-term photometry, high contrast imaging; see [talk by Courtney Dressing.](#))

**Through a cosmic convergence of resources, stellar astronomy will experience a renaissance** — stars will be real "stars" again, interesting for themselves, not just as planet hosts. (Needed capabilities include: interferometric imaging, speckle, high contrast imaging, long-term monitoring, precision photometry and spectroscopy, low- to moderate-resolution spectroscopy; see talks by [Gail Schaefer](#), [Jim Fuller](#), [David Nidever](#), [Juna Kollmeier](#), and [https://www.noao.edu/meetings/2020decadal/files/Bright-Universe-Wish-List.pdf](https://www.noao.edu/meetings/2020decadal/files/Bright-Universe-Wish-List.pdf))

**Time domain surveys and multi-messenger methods**, supported by ground-based observations, **will uncover new phenomena, some unimagined!** (see talks by Keith Bechtol, David Sand; Both computing and observing follow-up resources are needed on 1-30m aperture telescopes. The latter are often workhorse observing capabilities. Also required: alert brokers, flexible scheduling of follow-up resources, efficient data reduction protocols and pipelines; and protocols to quickly communicate results; see time domain breakout report [https://www.noao.edu/meetings/2020decadal/files/td-report.pdf](https://www.noao.edu/meetings/2020decadal/files/td-report.pdf))

**Spectroscopy will be the new imaging** — i.e., a widespread, multi-faceted discovery tool. The routine acquisition of millions of spectra through massively-multiplexed multi-object spectroscopy will open new doorways to discovery. (DESI on Mayall, PFS on Subaru in the first half of the coming decade. Future possibilities include MSE or a Future ESO facility; see WF Spectroscopy panel discussion and [https://www.noao.edu/meetings/2020decadal/files/noao_wide-field-spectroscopy-final.pdf](https://www.noao.edu/meetings/2020decadal/files/noao_wide-field-spectroscopy-final.pdf)).

**There will be many paths to success** — a more general message from the meeting is that astronomy in the 2020s will be driven by diverse questions, an abundance of new research tools, and openly available data. While future big-ticket items will continue to be important for discovery (JWST and its successors, ELTs), making advances and discoveries in astronomy will not be contingent on these: a multitude of existing facilities, capably instrumented, will open new horizons, solve problems, and raise new questions.

## 3. A Sustainable Future for Ground-based OIR Astronomy

The above diversity of research paths for the future is an important element to consider in Decadal Survey planning. It is also an important consideration in planning for the long-term future and health of OIR ground-based astronomy. Messages from the past few decades provide valuable context for the planning effort.

### Big Things Cost a Lot

One message from previous decades is that big facilities cost a lot and bigger facilities will cost more. In the late 1990's, HST and Keck were new, resounding successes, the stock market was climbing with irrational exuberance, and the astronomical community was exhorted to think bigger and imagine "faster, better, cheaper" paths to success. (http://www.spaceref.com/news/viewnews.html?id=864).

Some of the major projects from the 2000 Decadal Survey, JWST and GSMT, have been only partly successful in this regard. Both are more costly than originally estimated (JWST originally at $500M and now at $9B) and have taken longer to deliver than originally planned. Other high priorities have been or are being delivered, although their operations costs are daunting (LSST, ALMA). These results show that it is difficult to "break the cost curve", i.e., big projects really cost a lot to build and operate, and there is no free lunch.

In a recent *Nature* commentary, Matt Mountain and Adam Cohen have described how, as a result of the connection between size and cost, the US community faces a daunting task in the coming Decadal Survey. While each generation of facilities is more expensive to build and operate, the research budget of the National Science Foundation (NSF) has remained roughly flat over the past decade. In contrast, funding for astronomy in Europe is robust and stable, with the European Southern Observatory (ESO) having fully funded its 39-m Extremely Large Telescope (ELT).

Mountain and Cohen argue that ever-larger facilities are critical to exploration, and that without increased funding and longer-term planning, the US will cede its leadership in astronomy to Europe. An expansion of funding is required because merely shuttering or divesting from smaller facilities cannot meet even the operational costs of new facilities.

### A Youthful Scientific Field

The full picture is not as bleak (or unidirectional) as the one painted above when we consider the discovery track record in astronomy. The multitude of exciting discoveries from the last few decades illustrates how astronomy remains a youthful field with undiminished potential to surprise us and change the way we think about the Universe. In contrast to fields of research that may be dominated by a few important questions or quests, astronomy reaches out in numerous directions simultaneously.

As a result, discoveries are made in diverse ways. Some result from deliberate searches, while others arise completely unexpectedly. Some may be driven by technological developments, others by human inspiration and/or dogged persistence. Some are made by large projects and large organized teams, while others are made by small groups or individuals. The observing resources used to make discoveries range from relatively modest to state-of-the-art. Table 3 lists some recent major astronomical discoveries of the past few decades that have made use of ground-based OIR resources.

**Table 3. Some Major Astronomical Discoveries**

|  | Epoch of discovery | Discovery Resource |
|---|---|---|
| Discovery of the Kuiper Belt* | Jewitt & Luu 1993 | UH 88" |
| Exoplanets discovered* | 1995; Mayor & Queloz (Spitzer 2005 light from exoplanet; 2007 molecules in atm; 2009 weather map) | Mayor & Queloz: <br><br>Marcy & Butler: 3m Shane and 0.6m CAT. |
| Black hole at the center of the Milky Way* and tests of GR. | 1996-present | 8m VLT, 10m Keck |
| Dark Energy discovered* | 1998 | 2-4m discovery imaging; spectroscopy at Keck, ESO; follow up imaging with HST, CTIO, WIYN, INT, ESO |
| Dwarf planets in the outer Solar System* | 2002 Quaoar; 2005 Eris; 2016 evidence for Planet X | Quaoar Palomar 48"; 4m Blanco/DECam, 8m Subaru/HSC |
| Measurement of baryon acoustic oscillations, a new cosmological tool* | Cole et al. 2005; Eisenstein et al. 2005 | 4m AAO/2dF; 2.5m SDSS |
| Properties and occurrence rates of planets – e.g., super-Earths and Neptunes are common companions | Microlensing, e.g., Cassan et al. 2012 | 1.3m OGLE, 1m PLANET |
| Milky Way companions and merger history of the Milky Way: stellar streams and dwarf galaxies | 1994 Sgr dwarf, etc. | 1.3m 2MASS, 2.5m SDSS, 4m Blanco/DECam |
| Galaxies and quasars beyond z=7, patchiness of reionization | 2011+ | 8-10m spectroscopy, HST, Spitzer; 4m imaging, WISE |
| Optical counterpart to a binary neutron star merger and origin of rare heavy elements | 2017 | Small to large aperture |
| 'Oumuamua: a visiting planetesimal from another Solar System | 2017 | 1.8m PanSTARRS |

* Associated with one or more of these prizes: Bruce Medal, National Medal of Science, Nobel, Shaw Prize in Astronomy, Crafoord Prize, Kavli Prize, Breakthrough Prize.

The chart shows that facilities with apertures ranging from small to large have contributed to major discoveries in earlier decades, as has the opening of new frontiers such as the time domain. The diversity of paths to discovery has been supported by the traditional model of a diverse suite of ground-based OIR telescopes and instruments, both public and private that, taken as a whole, is accessible to a broad community of astronomers, each with their own approach to discovery.

While it is well known that smaller aperture facilities often support the research conducted in larger aperture facilities (e.g., imaging on smaller aperture facilities, spectroscopy on larger aperture facilities), the chart also illustrates how, perhaps surprisingly, relatively small aperture facilities have played the leading role in many major discoveries, even when larger aperture facilities are in existence. Perhaps this outcome is a result of the ability to carry out longer-term programs on smaller facilities and to take bigger risks in choosing which scientific challenges to take on.

## Many Revolutions Now

Based on the discovery track record in astronomy, modest-aperture facilities offer attractive growth options, especially when recycling is in the picture: many smaller revolutions can happen quickly and nimbly with modest investments, in contrast to the long development times of major facilities.

As described in the NOAO Decadal Survey planning meeting and the discoveries of previous decades, new frontiers such as the time domain (via a host of experiments) and the outer solar system are being explored on the ground with relatively small telescopes.

Equipping older facilities with new instrumentation and repurposing them for new missions (e.g., wide-field imaging such as DECam on 4m Blanco; highly-multiplexed MOS, as in DESI on 4m Mayall, PSF on 8m Subaru; NEID – high precision RV spectroscopy on 4m WIYN; ZTF at Palomar 48"), a trend this decade, is a strategy that can cut cost and development time. DESI was conceived in early 2009 and will begin commissioning in 2019. The extreme precision RV spectrograph NEID, which builds on state-of-the-art instrumentation experience, has a mere few-year development timeline.

With greater multiplexing (wider fields of view, larger format detectors, greater wavelength coverage, more spectra simultaneously), another time-honored approach, a smaller aperture facility can perform like a larger aperture facility, at modest cost.

## A Sustainable Future for Ground-based Astronomy

The above issues highlight an important issue for the coming Decadal Survey: given the likely constraints on astronomy funding in the coming decades, what is the best mix of investments in large, medium, and small capabilities that will ensure a bright future for astronomical discovery and put us on a path to a sustainable future? Wise investments will not only lead to discoveries with the facilities they fund, but also develop the tools and technologies that allow us to take the *next* step to future facilities and capabilities.

For example, mastering extreme precision radial velocities on small apertures makes it easy to extend that capability to larger apertures. Similarly for massively multiplexed spectroscopy. Sky surveys, both current and forthcoming (e.g., PanSTARRS, ZTF, DES, LSST), are teaching us a new skill: how to sift and serve massive amounts of data, a skill that may lead to future initiatives.

When we build extremely large telescopes, what technologies and techniques are we mastering that make successor initiatives relatively easy to carry out? If the answer is "to build an even larger telescope" and > 30m telescopes cannot be built because they cost too much, that investment, if it is the only one we make, will have taken us toward an unsustainable future, an evolutionary dead end.

More likely, a balanced set of investments in small to large categories is essential to a sustainable future. Based on the experience of previous decades, large facilities are the "value" investments that are guaranteed to produce compelling science and discoveries, while small and medium facilities are the "growth stocks" that are likely to deliver the biggest science bang per buck, sometimes with outsize returns (e.g., discovery of exoplanets, dark matter, dark energy). Investments in the latter category are critical because that is where growth arises reliably and at modest cost.

## Science, Workforce, and Society

Another important sustainability issue concerns our workforce. In previous decades, taking full advantage of the science opportunities offered by missions such as the NASA Great Observatories has required many students and postdocs. As a result, astronomy produced more highly trained students and postdocs than the available number of permanent positions in astronomy, a mismatch that has been a source of distress for our community.

Extracting the best science from future missions and facilities will likely continue to require the effort of students and postdocs. For researchers who work on data intensive projects like LSST, the data science skills that they acquire as part of their training will be directly relevant to many careers beyond astronomy – analytics, weather, climate, transportation, finance, marketing, and more – a situation that may partly resolve the mismatch and lead to a more sustainable workforce flow. By providing a highly trained workforce that is well suited to address important problems beyond astronomy, investments in data-intensive missions and their workforce also have additional benefits to society beyond the science they deliver.

These ideas find strong synergy in two of the [NSF's 10 Big Ideas](), namely *Harnessing the Data Revolution for 21st Century Science and Engineering* and *NSF INCLUDES*. NSF identifies *Harnessing the Data* as a frontier opportunity created by the confluence of "massive amounts of data online, [the ability to deploy] massive amounts of computation, and new algorithms for detecting hidden correlations and patterns in data" (see [NSF's video on its 10 big ideas](#)). One of the three components of this program is the "development of the 21st Century data-capable workforce". For example, a well-trained workforce can help "cities and workplaces become more livable through delivery of smart services informed by data, e.g., reduced highway and public transport congestion."

How do we engage people to join this workforce? How do we include diverse groups in this endeavor and in STEM careers in general, which is the aim of NSF INCLUDES? In the NSF video, France Cordova provides the answer, directly to the second question, but also implicitly to the first: "what's needed is access and inspiration." Translating this to the current discussion: training in data science, with application to real problems, can provide the access, while astronomy and its imagination-inspiring questions can provide the inspiration – a powerful combination!

To provide effective "access," now and into the future requires attention to data science trainers as much as to the trainees. With the rapid advances anticipated for computing and analysis techniques in the coming decade, it will be challenging for more senior scientists to keep pace with developing best practices and continue to fulfill their teaching and mentoring roles. Retraining and continuing education for senior scientists, as well as effective user support services, will likely be critical to support their efforts.

In short, astronomy offers the opportunity to solve real data science problems, start to finish, from conceptualization, to execution, analysis, and dissemination, at a high level of rigor. By training scientists who are well equipped to use their skills to solve problems in the public sector or private sector, astronomy can provide a valuable service to society by contributing to a data-capable workforce.

These benefits to society are arguably important considerations in prioritizing future astronomy-related investments. What weight should be given to these "broader impacts" in the prioritization process, e.g., compared to contributions to pure science knowledge? This is a question that the Decadal Survey should address.

## Summary

Looking at today's astronomy science and funding landscape, it is clear that we need a balanced set of investments in the small to large categories.

The "small" and "medium" categories are likely to deliver the biggest science "bang per buck," sometimes with outsize returns. They won't break the bank, and they can make valuable contributions beyond pure science, to society and our workforce. It is prudent to invest healthily in these categories, especially in sectors whose achievements open new evolutionary paths.

The "large" category is traditionally a value investment in that it is guaranteed to produce returns (discoveries). However, there is a risk associated with this category currently, both because of the science-to-cost ratio and whether the investment facilitates future evolution and growth. If there is adequate funding, investing in this category is certainly worthwhile.

## Appendix: NOAO Community Meeting Structure and Funding

Held 20-21 February 2018 in Tucson, AZ, the workshop included high-level science overviews of topics such as:
- Science with LSST and JWST
- Exoplanet Science in the 2020s
- The High Angular Resolution Universe
- Era of Bright (Star) Astronomy
- Galaxy Evolution and Stellar Populations
- Time Domain and Near-field Cosmology in the 2020s
- SDSS-V: Pioneering Panoptic Spectroscopy
- Science with GSMTs
- Maximizing the Science Return from Large Surveys
- Solar System Science in the Next Decade

and reviews of current implementation plans for needed capabilities, as well as break out sessions on science topics, techniques, and capabilities in need of development (GSMTs, Wide-field Spectroscopy, High Angular Resolution). The agenda at the meeting website includes links to the archived presentations and video streams of the plenary sessions (https://www.noao.edu/meetings/2020decadal/#agenda). The latter enabled remote participation during the workshop. The meeting was funded by NOAO and by the Kavli Foundation through the Kavli Futures Symposium program.